# Bipolar-doped superconducting infinite-layer cuprates

Fengzhe Wang[1†], Yueying Li[1,2†], Heng Wang[1,2†], Lizhi Xu[1†], Xianfeng Wu[1], Lixiang Xu[1], Guangdi Zhou[1,2], Jin-Feng Jia[1,2], Peng Li[1,2*], Haoliang Huang[1,2*], Qi-Kun Xue[1,2,3*], Zhuoyu Chen[1,2*]

[1]State Key Laboratory of Quantum Functional Materials, Department of Physics, Guangdong Provincial Key Laboratory of Topological Matter, and Guangdong Basic Research Center of Excellence for Quantum Science, Southern University of Science and Technology, Shenzhen 518055, China
[2]Quantum Science Center of Guangdong-Hong Kong-Macao Greater Bay Area, Shenzhen 518045, China
[3]Department of Physics, Tsinghua University, Beijing 100084, China

[†]These authors contributed equally.
*E-mail: chenzhuoyu@sustech.edu.cn, xueqk@sustech.edu.cn, huanghaoliang@quantumsc.cn, lipeng@quantumsc.cn

**Abstract**

Distilling the intrinsic physics of the superconducting $CuO_2$ plane from the complexities of charge-reservoir layers is a defining challenge in high-temperature superconductivity[1-3]. While superconducting electron-doped infinite-layer cuprates have been synthesized[4-9], controllable and uniform hole doping has long remained elusive despite exploratory attempts, limiting spectroscopic insights[10-14]. Here, we realize bipolar doping across infinite-layer $(Sr,Eu)CuO_2$ and $(Ca,Li)CuO_{2+\delta}$ single-crystalline thin films, mapping the electronic phase diagram. Both electron- and hole-doped films show pronounced electrical resistance anisotropy, indicating the quasi-two-dimensional nature of the $CuO_2$ planes. Angle-resolved photoemission spectroscopy across electron- and hole-doped regimes reveals persistent antiferromagnetic band folding coexisting with superconductivity. Remarkably, at a hole doping ~0.07 determined by Luttinger volume, the antiferromagnetic folding emerges from Fermi arcs within the film's single Fermi surface, with the onset superconducting transition temperature exceeding 60 K. These findings redefine the interplay between magnetic order and superconductivity and establish a definitive platform to investigate the intrinsic mechanism of high-temperature superconducting cuprates.

## Introduction

The discovery of high-temperature superconductivity in copper oxides in 1986 opened an extensive material family[15]. As superconducting cuprates were realized across hole- and electron-doped, single- and multilayer compounds, a common architecture became clear[1-3]: superconductivity resides in the $CuO_2$ planes, while the intervening charge-reservoir layers chiefly supply carriers. These reservoir layers, however, may also introduce structural complexity and competing degrees of freedom that obscure the intrinsic response of the $CuO_2$ plane. The infinite-layer structure[1], with $CuO_2$ planes stacked directly through alkaline-earth cations and without charge-reservoir layers, provides an ideal configuration to isolate the pristine physics of the superconducting planes. This structure is, however, thermodynamically metastable. Stabilizing this phase has been achieved through high-pressure bulk synthesis[4,10] and epitaxial thin-film growth. While steady progress has been achieved in electron-doped infinite-layer films[5-9], controllable and uniform hole doping within this pristine structure has remained difficult, precluding comprehensive spectroscopic characterization[11-14].

Progress in cuprate physics has advanced in step with spectroscopy techniques, and with angle-resolved photoemission spectroscopy (ARPES) in particular[16,17]. A robust dichotomy emerged between the two doping sides of the phase diagram. On the electron-doped side, antiferromagnetism and superconductivity coexist across a wide doping range, where the Fermi surface shows pronounced band folding across the antiferromagnetic zone boundary[18-21]. On the hole-doped side, the underdoped regime instead exhibits Fermi arcs that evolve into a full Fermi surface upon doping, with folding generally absent[22-25]. Recent ARPES investigations of hole-doped multilayer cuprates have revealed a more nuanced topology[26-28], in which inequivalent $CuO_2$ layers carry markedly different doping and disorder. The observed Fermi arcs are considered corresponding to the outer planes and small, antiferromagnetic-folded Fermi pockets are attributed to the more lightly doped inner planes. Coexistence of superconductivity and antiferromagnetism has been inferred for certain inner layers, but whether it is intrinsic or proximity-induced from neighbouring layers remains unresolved.

Here we access both electron and hole doping in infinite-layer cuprates by growing single-crystalline films by gigantic-oxidative atomic-layer-by-layer epitaxy (GAE), in which an exceptionally oxidative environment is maintained throughout the atomic-layer-by-layer epitaxial growth[29-32]. On the electron-doped side, we dope rare-earth element Eu in $SrCuO_2$ films on $GdScO_3$ substrates, and tune the carrier concentration through the Eu content. On the hole-doped side, we dope alkali metal Li in $CaCuO_{2+\delta}$ films on $NdGaO_3$ substrates, and tune the carrier concentration through the oxygen content (see methods). The resulting films present atomically flat surfaces that enable direct characterization by angle-resolved photoemission spectroscopy (ARPES). Because the hole concentration is set by a delicate oxygen content, we transfer the as-grown films from our home laboratory to the synchrotron in a cryogenic ultrahigh-vacuum suitcase, keeping them below 150 K throughout to preserve the surface oxygen stoichiometry[33,34]. Together, these capabilities deliver controllable and spectroscopy-ready electron- and hole-doped infinite-layer cuprates.

**Transport and bipolar phase diagram**

We construct a bipolar-doped phase diagram of the infinite-layer cuprate, in which both the electron- and hole-doped sides emanate from the common $3d^9$ insulating parent, using two distinct tuning routes (Fig. 1). On the electron-doped side, we control the nominal Eu concentration in $(Sr,Eu)CuO_2$ by varying the cation composition, and the normalized resistance–temperature (*R*–*T*) curves track the progressive establishment of superconductivity across this series (Fig. 1a and 1b left side). A complete, dome-shaped superconducting region is resolved, spanning Eu concentrations of roughly 0.05–0.17 and peaking near ~0.11, where the onset transition temperatures ($T_c^{onset}$, defined as the temperatures at which the resistance falls to 99.9% of the linearly extrapolated normal-state resistance) reaches about 30 K. On the hole-doped side, by contrast, the cation composition is fixed, and superconductivity is instead governed by the ozone partial pressure during growth (Fig. 1b right side and 1c). $T_c^{onset}$ rises monotonically with ozone pressure and reaches around 80 K at 0.06 mbar, already more than twice the maximum of the electron-doped side. This high $T_c^{onset}$ is achieved even though the hole doping remains within the underdoped regime, with the top of the dome not yet reached limited by present growth conditions. Such pronounced electron–hole asymmetry, captured within the infinite-layer structure, mirrors the long-known disparity between electron- and hole-doped cuprates[6,17,35-39]. On both sides, the normal state becomes increasingly metallic with rising carrier concentration.

To address the superconducting dimensionality, we probe the angular dependence of magnetoresistance for an electron-doped $Sr_{0.89}Eu_{0.11}CuO_2$ sample and a hole-doped sample (grown at 0.06 mbar ozone pressure, highest in the Fig. 1 phase diagram). For both, the *R*–*T* curves measured in out-of-plane and in-plane fields up to 14 T (Fig. 2a–d) display a pronounced anisotropy, the transition being suppressed far more strongly by out-of-plane fields. Extracting the upper critical field at 50% of the normal-state resistance, $B_c^{50\%}(T)$ in both cases can be fitted by the two-dimensional Ginzburg–Landau (GL) formula (Fig. 2e,g), yielding zero-temperature critical fields $B_{c0\perp}^{50\%} = 10$ T (19 T), $B_{c0\parallel}^{50\%} \approx 57$ T (116 T), an anisotropy $\gamma = B_{c0\parallel}^{50\%} / B_{c0\perp}^{50\%} \approx 5.7$ (6.0) and coherence lengths $\xi_{ab} \approx 5.7$ nm (4.2 nm) and $\xi_c \approx 1.0$ nm (0.7 nm), for the electron-doped (hole-doped) sample.

To distinguish two- from three-dimensional behaviour, we measure $R$–$T$ curves at various tilt angles $\theta$ in a fixed field (8 T for the electron-doped, 14 T for the hole-doped sample) and plot the temperature at 50% of the linear-extrapolated normal-state resistance, $T_c^{50\%}$, against $\theta$ (Fig. 2f,h). In both cases the data are captured by the Tinkham model for two-dimensional superconductors, whereas the anisotropic 3D-GL model fails to reproduce the sharp cusp near $\theta = 0°$. Both the electron- and hole-doped infinite-layer cuprates therefore exhibit pronounced quasi-two-dimensional superconductivity, indicating that the infinite-layer films closely approximate weakly coupled $CuO_2$ planes, making them a well-suited platform for probing the intrinsic physics of cuprate superconductivity.

**Fermi surfaces**

With the absence of charge-reservoir layers, ARPES on either the electron- or hole-doped films probes the $CuO_2$ plane directly. Varying the surface termination leaves the measured bands essentially unchanged (for instance, Ca terminated versus $CuO_2$ terminated in growth). Crucially, we resolve only a single Fermi surface in both doping cases, indicating a negligible doping gradient across the film thickness, in contrast to multilayer cuprates exhibiting multiple Fermi surfaces from inequivalent planes. This depth uniformity guarantees that these surface-sensitive measurements yield a Fermi surface representative of all $CuO_2$ planes .

Building on this, both doped regimes display the same essential feature: a single main Fermi surface coexisting with antiferromagnetic band folding, in superconducting samples. For the electron-doped (Sr,Eu)$CuO_2$ film ($T_c^{onset} = 26$ K), the Fermi surface map (Fig. 3a) and its tight-binding analysis (Fig. 3b) reveal a main band together with a folded band across the antiferromagnetic zone boundary (AFZB), corresponding to $n = 0.12$ electrons per Cu determined from the Luttinger volume, consistent with the nominal 11% substitution of Eu within error. The folded band near the antinodal region is directly visible in the energy–momentum spectra (Fig. 3e) and tracked by the momentum distribution curves (MDCs, Fig. 3f). The hole-doped (Ca,Li)$CuO_{2+\delta}$ film ($T_c^{onset} = 68$ K) markedly shows the folded bands near the nodal region (Fig. 3c,d), with the folding evident in the energy-momentum spectra and MDCs (Fig. 3g,h), corresponding to $p = 0.07$ holes per Cu from the Luttinger volume.

The hole-doped side is particularly striking for three reasons. First, the folded band sets in at the tips of the Fermi arcs, where the main band meets the AFZB, so that Fermi arcs and antiferromagnetic folding are observed together within the same Fermi surface. Second, this occurs at an exceptionally low hole doping, $p = 0.07$, which may explain why resistivity of the superconducting hole-doped samples are substantially higher.

Third, the lightly doped film sustains a relatively high superconducting transition temperature, with $T_c^{onset}$ exceeding 60 K.

These observations differ qualitatively from those in multilayer cuprates. In recent trilayer and quadruple-layer cuprate ARPES studies, the lightly doped inner $CuO_2$ planes seem to host high $T_c$ without antiferromagnetic folding[40,41]. For five-layer, six-layer and above, the lightly doped inner planes host small, antiferromagnetically folded electron pockets, while the outer planes show Fermi arcs[26-28]. Because these layers are highly different in doping and disorder, any coexistence of superconductivity and antiferromagnetism inferred for an inner plane is difficult to separate from interlayer proximity effects. Our films remove this ambiguity entirely: a single Fermi surface simultaneously exhibits both Fermi arcs and antiferromagnetic folding in a high-$T_c$ superconducting sample, establishing their coexistence as intrinsic.

**Structural Characterizations**

Figure 4 presents the structural characterizations of the infinite-layer cuprates. X-ray diffraction (XRD) θ–2θ scans (Fig. 4a) show that both the electron-doped $(Sr,Eu)CuO_2$ and hole-doped $(Ca,Li)CuO_{2+\delta}$ films are phase-pure, exhibiting exclusively (00*l*) Bragg reflections, with out-of-plane lattice constants of 3.384 Å and 3.189 Å, respectively. For $(Sr,Eu)CuO_2$ on $GdScO_3$, tensile strain compresses the *c* axis below the bulk $SrCuO_2$ value (~3.43 Å). For $(Ca,Li)CuO_{2+\delta}$ on lattice-matched $NdGaO_3$, the *c* axis is slightly reduced relative to bulk $CaCuO_2$ (~3.20 Å), consistent with the smaller ionic radius of Li substituting for Ca. The weaker $(Sr,Eu)CuO_2$ (001) reflection is attributed to its small structure factor. Pronounced Laue oscillations around the film peaks, together with reflection high-energy electron diffraction (RHEED) intensity oscillations during growth, confirm crystalline integrity and uniform thickness.

To resolve the atomic-scale structures, we performed cross-sectional scanning transmission electron microscopy (STEM) on a $NdGaO_3$/3-unit-cell (UC) $SrCuO_2$/13-UC $CaCuO_2$ heterostructure capped with $SrTiO_3$ (Fig. 4b,c). The $SrTiO_3$ cap is essential to shields the highly susceptible $CaCuO_2$ surface from Ga bombardment during focused-ion-beam preparation and absorbs part of the electron beam energy during imaging, preserving the intrinsic structure. All interfaces are atomically sharp, without discernible interdiffusion or misfit dislocations, indicating coherent epitaxy. Z-contrast in the high-angle annular dark-field (HAADF) image (Fig. 4b) cleanly distinguishes the $NdGaO_3$ substrate, $SrCuO_2$ buffer, $CaCuO_2$ film and $SrTiO_3$ cap, while the annular bright-field (ABF) image (Fig. 4c) resolves the oxygen columns and confirms the infinite-layer structure, with oxygen confined to the $CuO_2$ sheets.

Notably, the 3-UC $SrCuO_2$ buffer adopts an out-of-plane lattice constant far larger than the infinite-layer phase, matching instead the chain-type $SrCuO_2$ polymorph ($c \approx$ 3.91 Å). This structural selection is thickness-driven: below a critical thickness of ~4–5 UC, and under the compressive strain imposed by $NdGaO_3$ (~2% mismatch), the chain-type structure is favoured over the infinite-layer phase[42]. This buffer is, in turn, indispensable for layer-by-layer growth of $CaCuO_2$. As a highly metastable phase, $CaCuO_2$ is difficult to stabilize, and direct growth on $NdGaO_3$ triggers Stranski–Krastanov islanding driven by the structural and chemical mismatch with the perovskite surface. The $SrCuO_2$ buffer instead acts as a chemical template that lowers the $CaCuO_2$ nucleation barrier, converting the heteroepitaxy into a near-homoepitaxial process and restoring layer-by-layer growth. Although the buffer itself is chain-type, its top SrO planes reconstruct at the interface, enabling $CaCuO_2$ nucleation and coherent infinite-layer growth despite the underlying polymorphism.

**Summary**

In summary, the realization of bipolar-doped superconducting infinite-layer cuprates settles a long-standing materials bottleneck. By isolating the active $CuO_2$ planes from the structural and chemical complexities of charge reservoirs, these findings reconcile the historical electron-hole dichotomy regarding antiferromagnetism-superconductivity interplay within a single, intrinsic Fermi surface. The coexistence of prominent Fermi arcs and antiferromagnetic band folding at ultra-low hole doping challenges prevailing understanding, highlighting that magnetic order is intertwined with superconductivity at the microscopic level. This pristine material platform paves a definitive pathway toward decoding the fundamental pairing mechanism of high-temperature superconductivity.

**Methods**

**Thin film growth.** All the samples were grown using the gigantic-oxidative atomic layer-by-layer epitaxy (GAE) method[29] with real time monitoring via RHEED with electron energy of 30 keV.

The electron-doped $(Sr,Eu)CuO_2$ samples were grown on the (001)-oriented $GdScO_3$ substrates. Before film deposition, the $GdScO_3$ substrates are annealing at 1120°C for 2 h to obtain the GdO-terminated atomic surface[43]. During growth, the $Sr_{1-x}Eu_xO_y$ and $Cu_2O$ targets were alternately ablated using a KrF excimer laser with wavelength of 248 nm, where the cation stoichiometry can be precisely controlled by the exact laser pulse number. The calibration of the stoichiometry relies on the *in situ* RHEED intensity oscillations. The $Sr_{1-x}Eu_xO_y$ targets were synthesized from a stoichiometric mixture of $SrCO_3$ and $Eu_2O_3$ powders and sintered twice, each time for 10 h at 1100 °C. The $Cu_2O$ target is commercially available. The electron-doped $(Sr,Eu)CuO_2$ film growth was carried out at temperatures ranging from 660 to 920°C under a purified oxygen pressure of 0.01 mbar. The substrate temperature is measured by an infrared pyrometer from the back side of the Inconel600 sample holder, typically 70-80 °C higher than the temperature measured from the sample surface. The laser fluence was set to 1.2-1.5 $J\cdot cm^{-2}$, and the pulsed laser repetition rate was 3 Hz. Right after deposition, the samples were *in situ* annealed via $H_2$ with the pressure of $5 \times 10^{-4}$ mbar at 740°C for 20 min to remove the apical oxygen. Then the samples were cooled at a rate of $100°C\cdot min^{-1}$ to below 200°C before being transferred to the load-lock chamber.

The hole-doped $(Ca,Li)CuO_{2+\delta}$ samples were grown on the (001)-oriented $NdGaO_3$ substrates. Before film deposition, the $NdGaO_3$ substrates are annealing at 1150°C for 2 h to obtain the NdO-terminated atomic surface[43]. During growth, the $Ca_{0.5}Li_{0.5}O_y$ and $Cu_2O$ targets were alternately ablated and the calibration of the pulse number was also calibrated by the RHEED oscillations. The $Ca_{0.5}Li_{0.5}O_y$ target was synthesized from a stoichiometric mixture of CaO and $Li_2O$ powders and sintered twice, each time for 10 h at 800 °C. The hole-doped $(Ca,Li)CuO_{2+\delta}$ film growth was carried out at temperatures ranging from 660 to 850°C under purified ozone with different ozone pressure to adjust the oxygen content in the films. Right after deposition, the samples were cooled at a rate of $100°C\cdot min^{-1}$ to below 350°C before being transferred to the load-lock chamber.

**Transport and mutual inductance measurements.** Platinum (Pt) Hall-bar contacts were fabricated via magnetron sputtering. Electrical connections to the sample holders were subsequently established utilizing an ultrasonic aluminum wire bonder. All transport data were collected using a closed-cycle helium-free system (Physike) or a Physical Property Measurement System (Quantum Design). To ensure accurate in-plane alignment of the applied magnetic field, we calibrated the orientation by sweeping the rotation angle to locate the resistance minimum under a steady field and constant temperature. Finally, mutual inductance evaluations were conducted following the identical experimental protocol reported previously[30]. The upper critical fields were extracted from the temperature-dependent resistance curves measured at various magnetic fields, defined by the temperatures at which the resistance reaches 50% of the normal-state resistance. The in-plane ($\xi_{ab}$) and out-of-plane ($\xi_c$) coherence lengths were estimated within the anisotropic Ginzburg–Landau frameworks using $B_{c0\perp} = \phi_0/2\pi\xi_{ab}^2$ and $B_{c0//} = \phi_0/2\pi\xi_{ab}\xi_c$, where $\phi_0$ is the magnetic flux quantum. The superconducting anisotropy $\gamma$ is defined by the ratio of anisotropic coherence length or of penetration depth as $\gamma = \xi_{ab}/\xi_c$ (ref.[44]).

**ARPES measurements.** ARPES experiments were carried out at the BL03U

beamline of the Shanghai Synchrotron Radiation Facility (SSRF) in China. The system provided an energy resolution of better than 10 meV at 100-eV photons under an ultra-high vacuum (UHV) environment (base pressure below $7\times10^{-11}$ Torr). Following the film growth, the samples were transferred using distinct UHV suitcases depending on their doping type to prevent surface degradation. Specifically, electron-doped films were transferred in a room-temperature suitcase, whereas a cryogenic UHV suitcase (base pressure $< 5\times10^{-10}$ Torr, base temperature ~100 K) was deployed for the hole-doped films. This cryogenic setup enabled the rapid cooling of hole-doped samples to below 150 K. To completely suppress surface oxygen depletion and capture the intrinsic superconducting properties, the samples were strictly maintained at sub-150 K temperatures throughout the entire transit from the growth chamber to the ARPES endstation. Finally, all ARPES spectra were acquired with the sample temperature stabilized below 15 K.

**XRD and STEM measurements.** X-ray diffraction (XRD) $\theta$-$2\theta$ symmetric scans were performed using a Rigaku SmartLab X-ray diffractometer ($\lambda$=1.5406 Å). Specimens for cross-sectional STEM were prepared employing a FEI Helios 600i dual-beam focused ion beam (FIB) and scanning electron microscope machine. Prior to the extraction and thinning, protective platinum and carbon layer were deposited to prevent ion beam damage. STEM-HAADF and ABF images was acquired using a FEI Titan Themis G2 operated at 300 kV, outfitted with a double spherical-aberration corrector, a high-brightness field-emission gun (X-FEG) and a monochromator. The specific STEM imaging parameters were set with a probe semiconvergence angle of 25 mrad, alongside inner and outer collection angles ($\beta_1$ and $\beta_2$) of 48 and 200 mrad.

## Acknowledgements

We acknowledge the discussions with Xuchun Ma and Junfeng He. This work is supported by the National Key Research and Development Program of China (2024YFA1408101), the Natural Science Foundation of China (92565303, 92265112, 12374455, 12504165, 12504161, 12504166, & 52388201), the Guangdong Major Project of Basic Research (No. 2025B0303000004), the Guangdong Provincial Quantum Science Strategic Initiative (GDZX2501001, GDZX2401004 & GDZX2201001), the Shenzhen Science and Technology Program (KQTD20240729102026004), and the Shenzhen Municipal Funding Co-Construction Program Project (SZZX2301004 & SZZX2401001), and the support from the Station of Quantum Materials.

## Author contributions

Q.K.X. and Z.C. supervised the project. Z.C. initiated the study and coordinated the research efforts. F.W. performed thin film growth with L.X.X. and G.Z.'s assistance, under H.H. and Z.C.'s supervision. H.W. performed low-temperature measurements and analysis with X.W.'s assistance. Y.L., L.X. performed ARPES measurements and analysis under P.L. and Z.C.'s supervision. H.H. performed STEM and XRD analysis with F.W's assistance. Q.K.X., Z.C., and J.F.J. acquired funding. All authors participated discussions. Z.C. wrote the manuscript with input from all other authors.

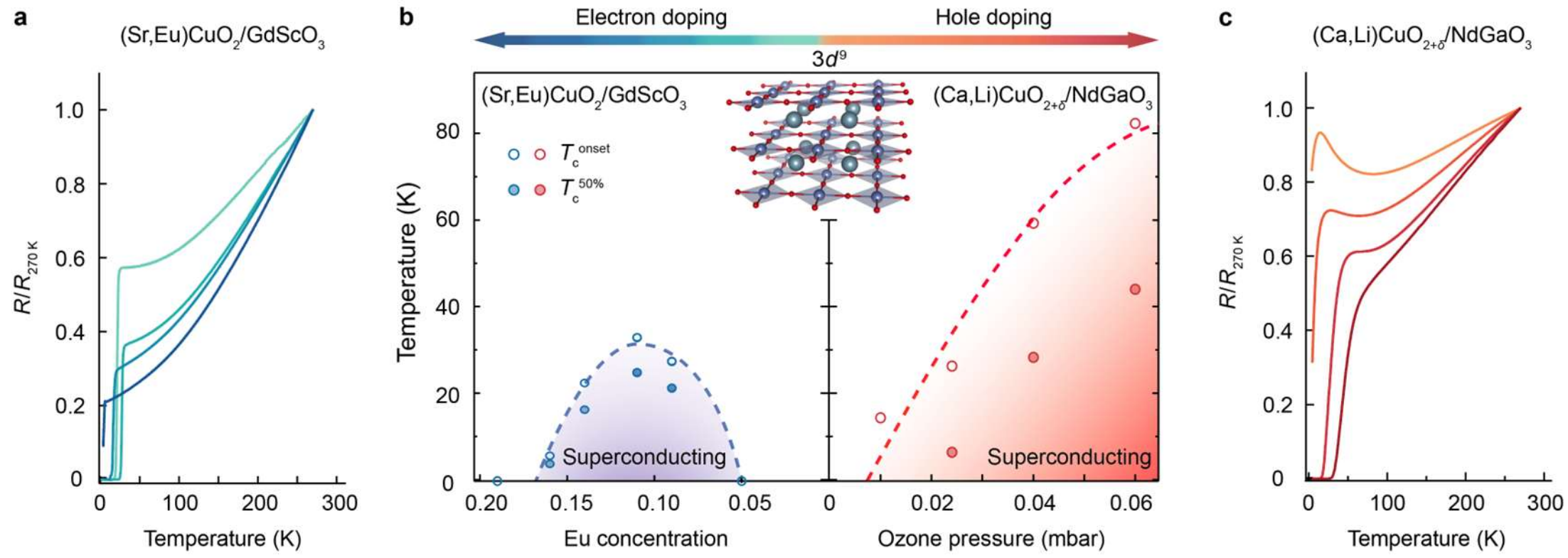

**Figure 1 | Phase diagram of bipolar-doped infinite-layer cuprates**. **a,** Normalized resistance-temperature curves (*R*–*T*) of representative electron-doped (Sr,Eu)$CuO_2$ samples with nominal Eu doping levels of 9%, 11%, 14% and 16%. **b,** the bipolar-doped phase diagram of infinite-layer cuprates. The inset shows the schematic crystal structure of infinite layer phase. Open and closed circles mark the $T_c^{onset}$ and $T_c^{50\%}$, which are defined as the temperatures at which the resistance deviated from the linear-fitted normal state resistance by 0.1% and 50%, respectively. **c,** The normalized *R*–*T* curves of representative hole-doped (Ca,Li)$CuO_{2+\delta}$ samples grown under different ozone pressures of 0.01, 0.024, 0.04 and 0.06 mbar.

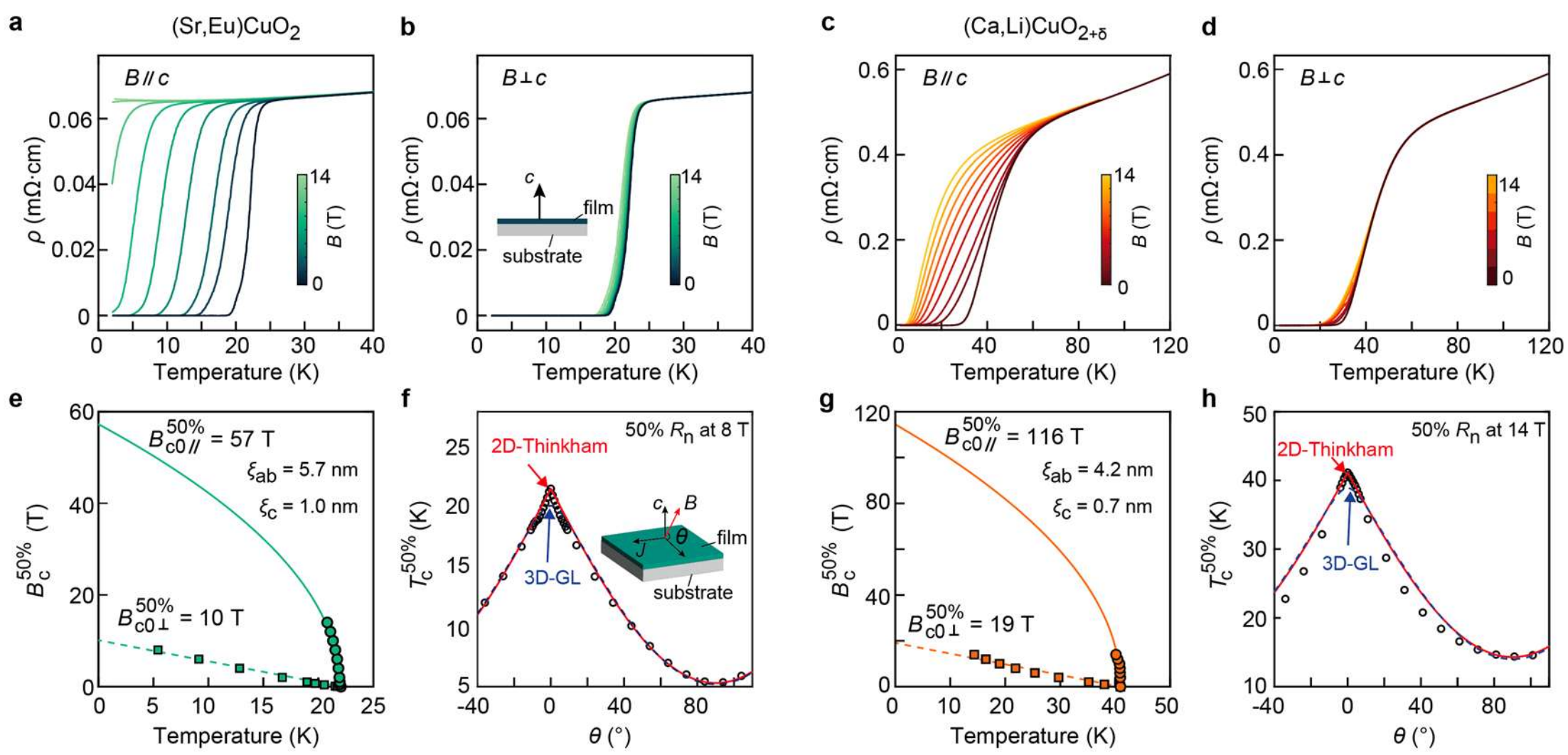

**Figure 2 | Quasi-two-dimensional superconductivity in both electron- and hole-doped infinite-layer cuprates**. **a,b**, $R$–$T$ curves of an electron-doped $Sr_{0.89}Eu_{0.11}CuO_2$ sample measured under out-of-plane (**a**) and in-plane (**b**) magnetic fields up to 14 T, revealing strong anisotropy. **c,d**, $R$–$T$ curves for a hole-doped sample (grown under 0.06 mbar $O_3$ partial pressure) under out-of-plane (**c**) and in-plane (**d**) fields. **e**, Temperature dependence of the upper critical field $B_c^{50\%}$ for the electron-doped sample, extracted at 50% of the normal-state resistance $R_n$, in-plane (solid circles) and out-of-plane (solid squares). The solid and dashed lines represent a fit to the 2D Ginzburg–Landau (GL) formula for in-plane and out-of-plane magnetic field, respectively. **f**, Angular dependence of $T_c^{50\%}$ (defined by 50% of $R_n$) measured at a fixed field of 8 T for the electron-doped sample. Inset shows the measurement geometry. **g**, $B_c^{50\%}(T)$ for the hole-doped sample. **h**, Angular dependence of $T_c^{50\%}$ at 14 T for the hole-doped sample. For both sides of doping, the Tinkham model for 2D superconductors (solid red lines in f and h) fits the data, whereas the anisotropic 3D GL model (dashed blue lines in f and h) fails to capture the cusp feature near $\theta = 0°$.

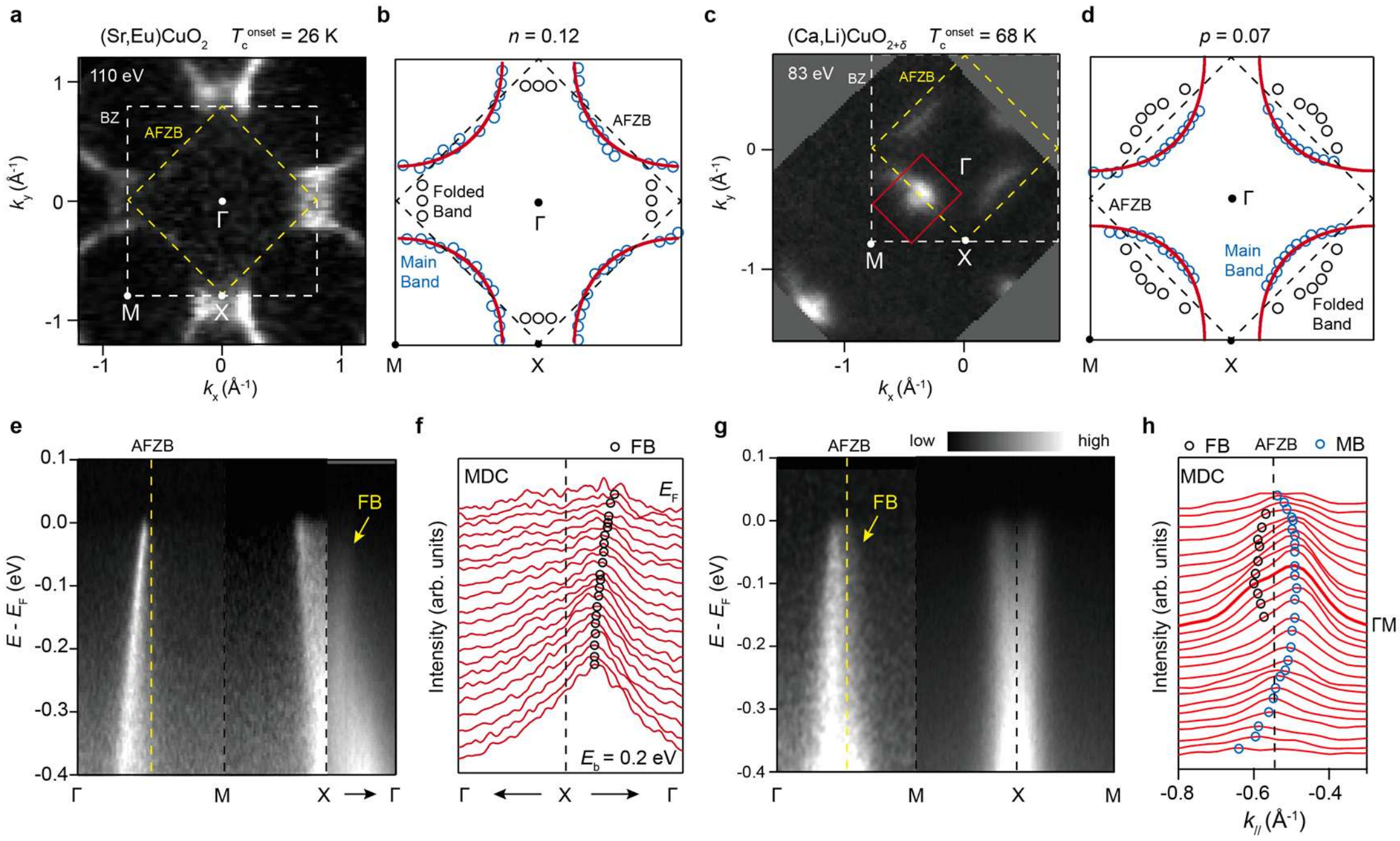

**Figure 3 | Electronic structure of electron- and hole-doped infinite-layer cuprate films. a,c,** Fermi surface maps measured using photon energy of 110 eV and 83 eV for $(Sr,Eu)CuO_2$ and $(Ca,Li)CuO_{2+\delta}$ films, respectively. White and yellow dashed lines denote the Brillouin Zone (BZ) boundaries and antiferromagnetic zone boundaries (AFZB), respectively. **b,d,** Tight-binding fitting of the Fermi surfaces indicates an electron doping of $n = 0.12$ electrons per Cu and a hole doping of $p = 0.07$ holes per Cu. Blue and black circles represent the intensity peak positions of the main bands and the folded bands, respectively. **e,g,** Energy-momentum spectra along high symmetry directions of electron- and hole-doped samples, respectively. Yellow arrows indicate the folded bands along representative directions. **f,** Momentum distribution curves (MDCs) of the folded band shown in panel **e**. **h,** MDCs showcasing the band folding within the red rectangular region in panel **c**. The gray-scale bar represents the ARPES intensity. FB: folded band. MB: main band.

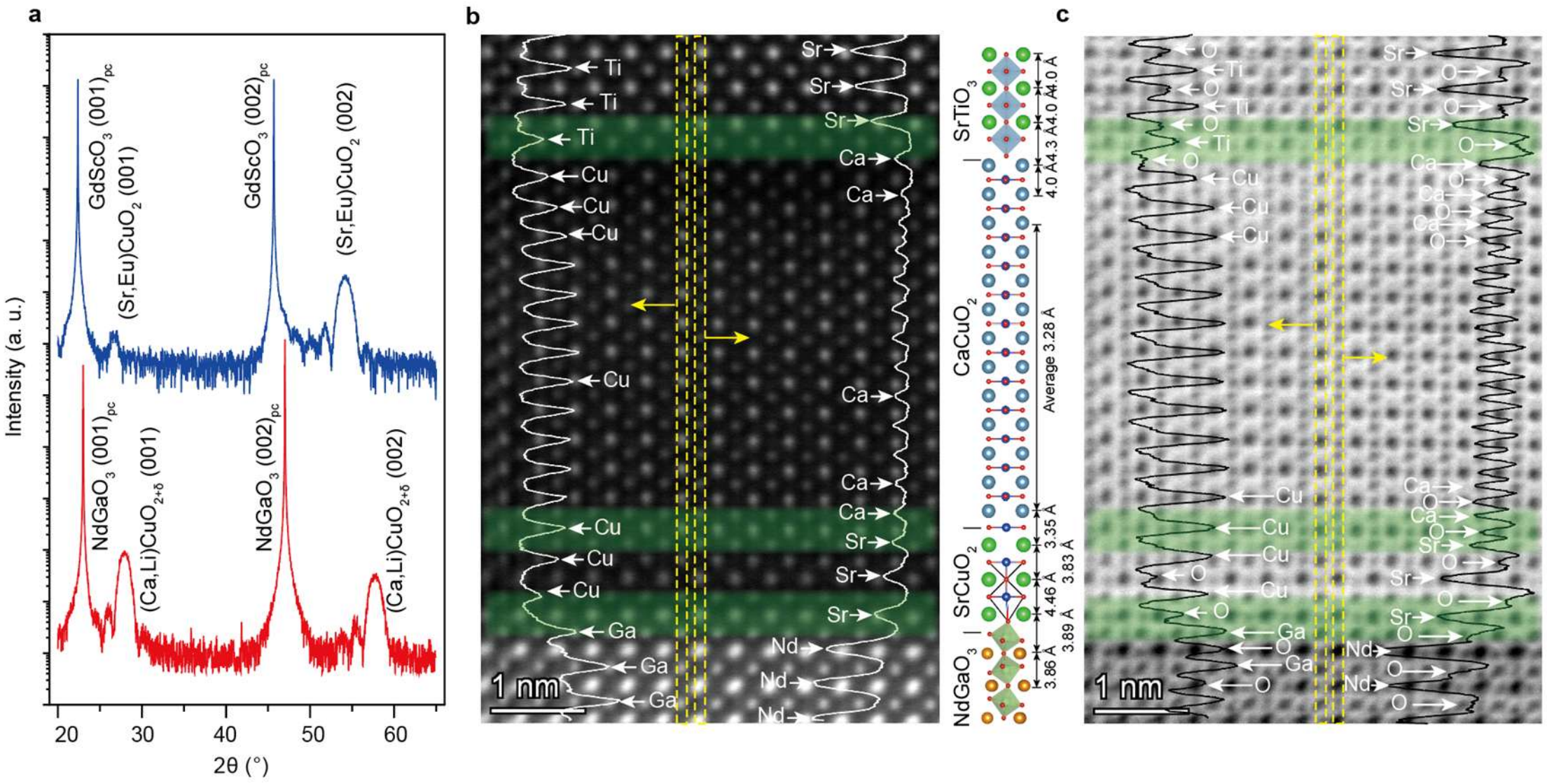

**Figure 4 | Structural characterization of infinite-layer cuprates.** **a**, X-ray diffraction θ–2θ scans of a c-axis-oriented $(Sr,Eu)CuO_2$ film grown on $GdScO_3$ substrate (top, blue) and a $(Ca,Li)CuO_{2+\delta}$ film grown on $NdGaO_3$ substrate (bottom, red). Only (00*l*) Bragg reflections are observed, confirming phase-pure epitaxial growth with [00*l*] orientation. **b**, Cross-sectional high-angle annular dark-field (HAADF) scanning transmission electron microscopy (STEM) image of a $NdGaO_3$/3-unit-cell (UC) $SrCuO_2$/13-UC $CaCuO_3$ stack capped with a $SrTiO_3$ capping layer, viewed along the $[100]_{pc}$ zone axis. Horizontal, green-shaded bands mark the heterointerfaces; vertical yellow dashed boxes indicate atomic columns from which intensity profiles are extracted and displayed on either side. Atomic columns are identified by Z-contrast: Nd (brightest), Sr, Ga, Cu, Ca and Ti. **c**, Corresponding annular bright-field (ABF) STEM image of the exact same region as panel **b**, resolving oxygen columns within the $CuO_2$ planes and verifying the infinite-layer structure. Intensity profiles extracted from the boxed columns confirm the oxygen sublattice. The atomic schematic between **b** and **c** illustrates the cation coordination and epitaxial relations across the perovskite substrate, buffer layer, infinite-layer film and perovskite cap.